\documentclass[
    ,final            
  ]
  {aipproc}

\layoutstyle{6x9}

\usepackage{color}
\usepackage{colordvi}

\graphicspath{{Pics/}}
\DeclareGraphicsExtensions{.eps,.ps}

\newcommand{\gsim}{\raisebox{-0.07cm}{$\, \stackrel{>}{{\scriptstyle
\sim}}\, $}}

\begin{document}
\vspace*{-25mm}
{\small DESY 11-104 \hfill DO-TH 11/17 \hfill LPN 11-35 \hfill SFB/CPP-11-34 \hfill TTK--11--21 
}
\vspace*{5mm}

\title{3-Loop Heavy Flavor Corrections to DIS with two Massive Fermion Lines}

\classification{11.10.Hi, 12.38.Bx, 12.38.Qk, 12.39.St, 14.65.Dw, 14.65.Fy}
\keywords      {High-energy scattering, parton distribution functions, heavy flavor corrections}

\author{J.~Ablinger}{
  address={
RISC, Johannes Kepler Universit\"at Linz, Altenberger Str. 69, A-4040 Linz, Austria 
}
}

\author{J.~Bl\"umlein}{
 address={Deutsches Elektronen-Synchrotron DESY, 
    Platanenallee 6, D--15738 Zeuthen, Germany
}
}
\author{S. Klein}{
 address={
Institute f\"ur Theoretische Teilchenphysik und Kosmologie, RWTH Aachen University, \\  D-52056 Aachen, 
Germany
}
}

\author{C. Schneider}{
 address={
RISC, Johannes Kepler Universit\"at Linz, Altenberger Str. 69, A-4040 Linz, Austria 
}
}
\author{F. Wi\ss{}brock}{
 address={Deutsches Elektronen-Synchrotron DESY, 
    Platanenallee 6, D--15738 Zeuthen, Germany
}
}

\begin{abstract}
We report on recent results obtained for the massive operator matrix elements which contribute
to the massive Wilson coefficients in deep-inelastic scattering for $Q^2 \gg m_i^2$ in case of
sub-processes with two fermion lines and different mass assignment.
\end{abstract}

\maketitle
\newcommand\N{\nonumber}
\newcommand\ep{\varepsilon}
\newcommand{\Ahathat}{\hat{\hspace*{-1mm}\hat{A}}}

\section{Introduction}

\noindent
To determine the value of the strong coupling constant $\alpha_s(M_Z^2)$ from the world
deep-inelastic scattering data \cite{ALPH} precisely, the structure function $F_2(x,Q^2)$ has to be 
described to 
$O(\alpha_s^3)$ accuracy at present. The corresponding next-to-next-to-leading order (NNLO) parton 
distributions 
are
instrumental for accurate predictions of the weak boson and Higgs boson production cross sections
at hadron colliders \cite{Alekhin:2010dd}.
While the massless contributions are known to NNLO \cite{MVV}, in the 
massive case this has been achieved for a series of Mellin moments $N$ only in the region $Q^2 \gsim 10 
m_Q^2$ \cite{BBN}~\footnote{The $O(\alpha_s^2)$ corrections were given in \cite{NLO}; see also the 
numerical implementation in Mellin space \cite{AB}.}, with $m_Q$ the mass of the heavy quark. Based 
on the 2-loop massive operator matrix elements, which are known for general values of $N$  
\cite{TL1} to the linear term in the dimensional parameter $\varepsilon$ \cite{TL2}, and the massless 
3-loop Wilson coefficients \cite{MVV1} one may compute all logarithmic contributions at 3--loop order 
\cite{LOG1}. In the kinematic region of HERA, however, these terms alone are not dominant 
over the yet unknown constant term for general values of $N$, as has been demonstrated for the moments 
calculated in \cite{BBN}. The $O(\alpha_s^3)$ corrections to the structure function $F_L(x,Q^2)$ in the 
asymptotic region were calculated in \cite{FL,BBN}. Heavy flavor corrections for charged current 
reactions were given in \cite{CC}.

The heavy flavor contributions to the structure function $F_2(x,Q^2)$ at $O(\alpha_s^3)$ in case of one 
massive quark are described by the five massive Wilson coefficients in the asymptotic region $L_q^{\rm 
NS}, L_q^{\rm PS},
L_g^{\rm S}, H_q^{\rm PS}, H_g^{\rm S}$ \cite{BBN}. Two of these Wilson 
coefficients,  $L_q^{\rm PS}$ and $L_g^{\rm S}$, have been computed completely for general values of 
$N$ in \cite{ABKSW}, cf. also \cite{LOG1}. In \cite{ABKSW} the contributions to the color 
factors $O(T_F^2 N_f C_{A,F})$ of the Wilson coefficients $L_q^{\rm NS}, H_q^{\rm PS}, H_g^{\rm S}$ 
were also calculated. The corresponding Feynman diagrams consist of graphs with two internal fermion 
lines, out of which one is massless and one massive. After applying algebraic relations 
\cite{ALGEBRA} these contributions to the massive Wilson coefficients can be represented in terms of 
the known set of weight {\sf w = 4}  harmonic sums \cite{BRK}.

Related cases concern the $O(\alpha_s^3)$ contributions with two massive lines $\propto T_F^2 C_{F,A}$
with either equal or different masses. In this note we present first results for these contributions.
\section{Two massive quarks of equal mass}

\noindent
We first consider the case of two quarks of equal mass in the $O(\alpha_s^3)$ operator matrix elements
$\propto T_F^2$. In calculating the corresponding Feynman diagrams we apply the Mellin-Barnes
representation \cite{PK} through which the Feynman parameter integrals can be expressed in terms
of Meijer $G$-functions \cite{MEIJ} in general. In these calculations, like those in \cite{ABKSW},
the use of modern summation algorithms as encoded in {\tt Sigma} \cite{SIGMA} are of essential 
importance. For the quarkonic flavor non-singlet 
case one obtains for the constant contribution to the operator matrix element~:

{\footnotesize
\begin{eqnarray} 
\hat{a}_{qq,Q}^{(3),NS}(N)&=&
    {T_F^2 C_F} \Black{}\Biggl\{
\frac{128}{27} {S_4} \Black{}
-\frac{1024 }{27} \zeta_3 S_1
+\frac{64}{9} \zeta_2 S_2
+\frac{256  \left(3 N^2+3 N+2\right) }{27 N (N+1)} \zeta_3
-\frac{320 }{27} \zeta_2 S_1
-\frac{640  }{81} S_3
\N\\&&
 +\frac{8  \left(3 N^4+6 N^3+47 N^2+20 N-12\right) }{27 N^2 (N+1)^2} \zeta_2
+\frac{1856 }{81} S_2
-\frac{19424 }{729} S_1
 -\frac{4  P_1(N)
}{729 N^4 
(N+1)^4}
        \Biggr\}.
\end{eqnarray}
}
 
\noindent
The results in the flavor pure-singlet case are~:

{\footnotesize
\begin{eqnarray}
\hat{a}^{(3), \rm PS}_{Qq}(N) &=&
\frac{T_F^2 C_F}{(N-1)N^2(N+1)^2(N+2)}
\Biggl\{
(N^2+N+2)^2 \Biggl[\frac{32}{27} S_1^3 - \frac{512}{27} S_3 + \frac{128}{3} S_{2,1}
-\frac{1024}{9} \zeta_3 
\nonumber\\ 
&&
\hspace*{-20mm}
- \frac{160}{9} S_2 S_1 + \frac{32}{3} \zeta_2 S_1 \Biggr]
- \frac{32 P_3(N) \zeta_2}{9N(N+2)} 
+ \frac{32 P_4(N) S_2}{27 N(N+2)(N+3)(N+4)(N+5)}
\nonumber\\ &&
\hspace*{-20mm}
- \frac{32 P_5(N) S_1^2}{27 N(N+1)(N+2)(N+3)(N+4)(N+5)}
+ \frac{64 P_6(N) S_1}{81 N^2(N+1)^2(N+2)^2(N+3)(N+4)(N+5)}
\nonumber\\ &&
\hspace*{-20mm}
- \frac{64 P_7(N) S_1}{243 N^3(N+1)^2(N+2)^3(N+3)(N+4)(N+5)}
\Biggr\}.
\end{eqnarray}
}

\noindent
Here $P_k(N)$ denote certain polynomials in $N$ and $S_{\vec{a}} \equiv S_{\vec{a}}(N)$ are harmonic
sums \cite{HSUM}.
These relations generalize the moments being obtained previously in \cite{BBN}. The  
contributions to the anomalous dimensions given in \cite{MVV,ANDIM2a} are confirmed. 
\section{Two massive quarks of unequal mass}

\noindent
Beginning with  $O(\alpha_s^3)$ graphs with internal massive fermion lines carrying unequal masses
contribute. Since the mass ratio in case of the charm and bottom quarks is given by $x =m_c^2/m_b^2 
\simeq 1/10$, one may expand the corresponding diagrams using this parameter. We first calculated
the 2nd and 4th moment of the gluonic operator matrix element $A_{Qg}$ extending the code {\tt qexp}
\cite{QEXP} to higher moments applying projectors similar to those used in \cite{BBN}. The 
unrenormalized operator matrix elements are given by~:

{\footnotesize
\begin{eqnarray}
& & \hspace*{-3mm}
\Ahathat_{Qg}^{(3)}(N=2) = \N\\
&& T_F^2 C_A \Biggl\{
          \frac {1} {\ep^3} \frac{8960}{81}
          +\frac {1} {\ep^2} \left[
          -\frac{1184}{243}
          + \frac{2240}{27}
           \left(\ln\left(\frac{m_1^2}{\mu^2}\right) 
          + \ln\left(\frac{m_2^2}{\mu^2}\right)\right) 
          \right]       
          +\frac {1} {\ep} \Biggl[
          \frac{35800}{729}
          + \frac{1120}{27} ~\zeta_2
\N\\&&
          + \frac{1168}{27}
          \left(
          \ln^2\left(\frac{m_1^2}{\mu^2}\right) 
          +\ln^2\left(\frac{m_2^2}{\mu^2}\right)\right) 
          -\frac{296}{81}
          \left(
          \ln\left(\frac{m_1^2}{\mu^2}\right) 
          +\ln\left(\frac{m_2^2}{\mu^2}\right)\right) 
          + \frac{1024}{27} \ln\left(\frac{m_1^2}{\mu^2}\right) \ln\left(\frac{m_2^2}{\mu^2}\right)~ 
          \Biggr]
\N\\&&
          + \frac{156458}{2187}
          - \frac{1696}{81} \zeta_3
          - \frac{148}{81} \zeta_2
          + \frac{512608}{10125} x
          + \frac{3130072}{496125} x^2
          + \frac{112173472}{843908625} x^3
          + \frac{18710}{243} \ln\left(\frac{m_1^2}{\mu^2}\right) 
\N\\&&
          - \frac{10}{3} \ln\left(\frac{m_2^2}{\mu^2}\right) 
          + \frac{280}{9} \zeta_2\left(
          \ln\left(\frac{m_1^2}{\mu^2}\right) 
          +\ln\left(\frac{m_2^2}{\mu^2}\right)\right) 
          + \Biggl[
           \frac{14368}{675} x
          + \frac{12016}{4725} x^2
          - \frac{328928}{2679075} x^3
          \Biggr] \ln\left(x\right)
\N\\ && 
          + \Biggl[
          - \frac{70}{81}
          - \frac{16}{45} x
          - \frac{16}{45} x^2
          - \frac{5104}{8505} x^3
          \Biggr] 
          \Biggl(\ln^2\left(\frac{m_1^2}{\mu^2}\right) 
          +\ln^2\left(\frac{m_2^2}{\mu^2}\right)\Biggr)
          + \Biggl[
          - \frac{304}{81}
          + \frac{32}{45} x
          + \frac{32}{45} x^2
          + \frac{10208}{8505} x^3
          \Biggr]
\N\\ &&
          \times \ln\left(\frac{m_1^2}{\mu^2}\right) \ln\left(\frac{m_2^2}{\mu^2}\right) 
          + \frac{208}{27}
          \ln^2\left(\frac{m_1^2}{\mu^2}\right) 
          \ln\left(\frac{m_2^2}{\mu^2}\right) 
          + \frac{560}{27}
          \ln\left(\frac{m_1^2}{\mu^2}\right) 
          \ln^2\left(\frac{m_2^2}{\mu^2}\right) 
          + \frac{1544}{81} \ln^3\left(\frac{m_1^2}{\mu^2}\right) 
\N\\ &&
          + \frac{1192}{81} \ln^3\left(\frac{m_2^2}{\mu^2}\right) 
          \Biggr\}
+ T_F^2 C_F \Biggl\{
          - \frac {1} {\ep^3} \frac{4096}{81}
          + \frac {1} {\ep^2} \Biggl[
          \frac{5120}{81}
          - \frac{1024}{27} \left(
          \ln\left(\frac{m_2^2}{\mu^2}\right) 
          + \ln\left(\frac{m_1^2}{\mu^2}\right)\right)
          \Biggr]
\N\\ &&
          + \frac {1} {\ep} \Biggl[
          \frac{28544}{729}
          - \frac{512}{27} \zeta_2 
          -\frac{256}{9}\Biggl( 
          \ln^2\left(\frac{m_2^2}{\mu^2}\right)
          + \ln^2\left(\frac{m_2^2}{\mu^2}\right)\Biggr)
          + \frac{1280}{27} \left(
          \ln\left(\frac{m_2^2}{\mu^2}\right)
          +\ln\left(\frac{m_2^2}{\mu^2}\right)\right)
          \Biggr]
          + \frac{128}{243}
\N\\&&
          + \frac{3584}{81} \zeta_3
          + \frac{640}{27} \zeta_2
          - \frac{1517888}{30375} x
          + \frac{339785728}{10418625} x^2
          + \frac{1653611968}{843908625} x^3
          +          \frac{256}{81}
          \ln\left(\frac{m_1^2}{\mu^2}\right) 
          + 
          \frac{13504}{243}
          \ln\left(\frac{m_2^2}{\mu^2}\right) 
\N\\ &&          
          - \frac{128}{9} \zeta_2
          \left(\ln\left(\frac{m_1^2}{\mu^2}\right) 
          +\ln\left(\frac{m_2^2}{\mu^2}\right)\right) 
+ \Biggl[
          - \frac{45952}{2025} x
          - \frac{2056384}{99225} x^2
          - \frac{11786368}{2679075} x^3
          \Biggr]
          \ln\left(x\right) 
          + \Biggl[
          \frac{1936}{81}
\N\\ &&
          + \frac{896}{135} x
          + \frac{9536}{945} x^2
          + \frac{47744}{8505} x^3
          \Biggr]
          \Biggl(\ln^2\left(\frac{m_1^2}{\mu^2}\right) 
          + \ln^2\left(\frac{m_2^2}{\mu^2}\right)\Biggr) 
          - \frac{256}{27}
          \ln\left(\frac{m_1^2}{\mu^2}\right) \ln\left(\frac{m_2^2}{\mu^2}\right) 
          \ln\left(x\right)
\N\\ &&
          + 
          \Biggl[
          \frac{1888}{81}
          - \frac{1792}{135} x
          - \frac{19072}{945} x^2
          - \frac{95488}{8505} x^3
          \Biggr]
          \ln\left(\frac{m_1^2}{\mu^2}\right)
          \ln\left(\frac{m_2^2}{\mu^2}\right)
          - \frac{1408}{81}
          \ln^3\left(\frac{m_1^2}{\mu^2}\right)
          - \frac{896}{81}
          \ln^3\left(\frac{m_2^2}{\mu^2}\right) 
          \Biggr\}
\nonumber\\ 
&& +O(x^4 \ln(x))~,
\label{eq:MOM2}
\end{eqnarray}
}

\noindent
where $m_1 < m_2, m_2^2 \ll \mu^2$. 
The 4th moment reads~:

\noindent
{\footnotesize
\begin{eqnarray}
& & \hspace*{-3mm}
\Ahathat_{Qg}^{(3)}(N=4) = \N\\
&&
T_F^2 C_A \Biggl\{
       \frac {1} {\ep^3}\frac{287408}{2025}
       +\frac {1} {\ep^2} \Biggl[
           \frac{11614}{135}
         + \frac{71852}{675}
                   \left( 
\ln\left(\frac{m_1^2}{\mu^2}\right)
+\ln\left(\frac{m_1^2}{\mu^2}\right)\right)
          \Biggr]
       +\frac {1} {\ep} \Biggl[
          \frac{264315863}{1822500}
 +\frac{35926}{675} \zeta_2
\N\\ &&         
+ \frac{12287}{225} \left(       
           \ln^2\left(\frac{m_1^2}{\mu^2}\right)
+           \ln^2\left(\frac{m_2^2}{\mu^2}\right)\right)
          +\frac{5807}{90}
\left(
\ln\left(\frac{m_1^2}{\mu^2}\right) 
+ \ln\left(\frac{m_1^2}{\mu^2}\right)\right) 
  +  \frac{3784}{75} \ln\left(\frac{m_1^2}{\mu^2}\right) \ln\left(\frac{m_2^2}{\mu^2}\right)
          \Biggr]
\N\\ &&        
          + \frac{4887988511}{24300000}
          - \frac{47146}{2025} \zeta_3
          + \frac{5807}{180} \zeta_2
          + \frac{496855133}{7441875} x
          + \frac{2510388298}{468838125} x^2
          + \frac{250077164867}{5616211899375} x^3
\N\\ &&
+           \frac{384762007}{2430000}
\ln\left(\frac{m_1^2}{\mu^2}\right) 
       + 
           \frac{47956573}{810000}
\ln\left(\frac{m_2^2}{\mu^2}\right) 
+\frac{17963}{450} \zeta_2
\left(
\ln\left(\frac{m_1^2}{\mu^2}\right) 
+\ln\left(\frac{m_2^2}{\mu^2}\right)\right) 
+ \Biggl[           
           \frac{1877399}{70875} x
\N\\ &&
          + \frac{3269548}{1488375} x^2
          - \frac{156082853}{1620840375} x^3  \Biggr]
\ln\left(x\right) 
+ \Biggl[
           \frac{532373}{16200}
          + \frac{707}{1350} x
          - \frac{284}{525} x^2
          - \frac{744283}{935550} x^3
          \Biggr]
          \Biggl( 
 \ln^2\left(\frac{m_1^2}{\mu^2}\right)
\N\\ &&
+ \ln^2\left(\frac{m_2^2}{\mu^2}\right)\Biggr)
+ \Biggl[ \frac{62893}{2025}
          - \frac{707}{675} x
          + \frac{568}{525} x^2
          + \frac{744283}{467775} x^3
          \Biggr]  \ln\left(\frac{m_1^2}{\mu^2}\right)  \ln\left(\frac{m_2^2}{\mu^2}\right)
          + \frac{7579}{675} \ln^2\left(\frac{m_1^2}{\mu^2}\right) 
\N\\ &&
\times \ln\left(\frac{m_2^2}{\mu^2}\right)   
          +  \frac{17963}{675}
\ln\left(\frac{m_1^2}{\mu^2}\right) \ln^2\left(\frac{m_2^2}{\mu^2}\right)   
          + \frac{3817}{162}   \ln^3\left(\frac{m_1^2}{\mu^2}\right)  
          + \frac{74657}{4050} \ln^3\left(\frac{m_2^2}{\mu^2}\right) 
          \Biggr\}
\N\\ &&
+ T_F^2 C_F \Biggl\{
          - \frac {1} {\ep^3} \frac{297616}{10125}
  + \frac {1} {\ep^2} \Biggl[
          \frac{2947282}{151875}
          - \frac{74404}{3375} \left(
  \ln\left(\frac{m_1^2}{\mu^2}\right)
+ \ln\left(\frac{m_2^2}{\mu^2}\right)\right)
          \Biggr]
          + \frac {1} {\ep} \Biggl[
           \frac{21662237}{9112500}
\N\\ &&
          - \frac{37202}{3375} \zeta_2
          - \frac{18601}{1125}
\left(\ln^2\left(\frac{m_1^2}{\mu^2}\right)
+\ln^2\left(\frac{m_2^2}{\mu^2}\right)\right)
                  + \frac{1473641}{101250}
\left(
\ln\left(\frac{m_1^2}{\mu^2}\right) 
+ \ln\left(\frac{m_2^2}{\mu^2}\right)\right) 
          \Biggr]
\N\\ &&
                    - \frac{33406758667}{1093500000}
          + \frac{260414}{10125} \zeta_3
          + \frac{1473641}{202500} \zeta_2
          - \frac{119314474}{4134375} x
          + \frac{582667691}{37507050} x^2
          + \frac{46049137562}{44929695195} x^3
\N\\ &&
            - \frac{37307959}{2430000}
\ln\left(\frac{m_1^2}{\mu^2}\right) 
          + \frac{76621423}{4050000}
\ln\left(\frac{m_2^2}{\mu^2}\right) 
           - \frac{18601}{2250} \zeta_2
\left(\ln\left(\frac{m_1^2}{\mu^2}\right) 
+ \ln\left(\frac{m_2^2}{\mu^2}\right)\right) 
+ \Biggl[
           \frac{368428}{39375} x
\N\\ &&
          + \frac{2876423}{297675} x^2
 + \frac{570093292}{324168075} x^3
          \Biggr]
 \ln\left(x\right) 
   + 
          \Biggl[
          \frac{530371}{81000}
          + \frac{4456}{1125} x
          + \frac{27101}{4725} x^2
          + \frac{1759616}{467775} x^3
          \Biggr]  
\N\\ &&
\times
\left(
\ln^2\left(\frac{m_1^2}{\mu^2}\right) 
+ \ln^2\left(\frac{m_2^2}{\mu^2}\right)\right) 
          + \frac{18601}{3375}
\ln\left(\frac{m_2^2}{\mu^2}\right) \ln\left(\frac{m_1^2}{\mu^2}\right) 
\ln\left(x\right) 
                    + 
 \Biggl[
                     \frac{442267}{50625}
- \frac{8912}{1125} x
\N\\ &&
          - \frac{54202}{4725} x^2
          - \frac{3519232}{467775} x^3
             \Biggr]
\ln\left(\frac{m_1^2}{\mu^2}\right) \ln\left(\frac{m_2^2}{\mu^2}\right)
          - \frac{204611}{20250} \ln^3\left(\frac{m_1^2}{\mu^2}\right) 
          - \frac{130207}{20250} \ln^3\left(\frac{m_2^2}{\mu^2}\right)
          \Biggr\}
\nonumber\\
&& +O(x^4 \ln(x))~.
\label{eq:MOM4}
\end{eqnarray}
}

\noindent
These contributions to the massive Wilson coefficients can be uniquely calculated in the fixed flavor 
number scheme with three massless quarks in the initial state. However, they cannot be attributed either
to the charm or bottom distribution in a variable flavor scheme.~\footnote{As has been shown in 
\cite{BVN} the application of a variable flavor scheme usually leaves no free choice of the matching 
scale $\mu^2$. In particular it is often quite different from the respective heavy flavor mass $m_Q^2$.}  
This shows one of the limitations
of this intention despite the fact that operator matrix elements are process independent 
quantities. In (\ref{eq:MOM2}, \ref{eq:MOM4}) power corrections contribute, however {\it not} of the 
kind $O(m_i^2/Q^2)$ but of $O(m_1^2/m_2^2)$. The computation of the corresponding contributions for 
general values of $N$ is in progress.

We remark that the representation of the heavy flavor Wilson coefficients to $O(\alpha_s^3)$
are given in the fixed flavor number scheme. As has been shown in Refs.~\cite{DO,ABKM} this 
choice is sufficient for the kinematic region at HERA.

This work has been supported in part by Studienstiftung des Deutschen Volkes, 
DFG Sonderforschungsbereich Transregio 9, 
Computergest\"utzte Theoretische
Teilchenphysik, Austrian Science Fund (FWF) grant P20347-N18,
and EU Network {\sf LHCPHENOnet} PITN-GA-2010-264564.

\bibliographystyle{aipproc}   
  
\bibliography{sample}


\end{document}